\documentclass{iau} 
\usepackage{graphicx}

\title[IAUS335. Flux Rope Input to CME forecasting ] 
{A new technique to provide realistic input to CME forecasting models}

\author[Nat Gopalswamy, Sachiko Akiyama, Seiji Yashiro, \& Hong Xie]   
{Nat Gopalswamy$^1$, Sachiko Akiyama$^2$, Seiji Yashiro$^2$,
 \and Hong Xie$^2$}

\affiliation{$^1$ Solar Physics Laboratory, NASA Goddard Space Flight Center, Greenbelt, MD 20771, USA
 \\ email: {\tt nat.gopalswamy@nasa.gov} \\[\affilskip]
$^2$Dept. of Physics, The Catholic University of America, Washington DC 20064, USA}

\pubyear{2017}
\volume{335}  
\jname{Space Weather of the Heliosphere: Processes and Forecasts}
\editors{C. Foullon \& O. Malandraki, eds.}
\begin{document}

\maketitle

\begin{abstract}
We report on a technique to construct a flux rope (FR) from eruption data at the Sun. The technique involves line-of-sight magnetic fields, post-eruption arcades in the corona, and white-light coronal mass ejections (CMEs) so that the FR geometric and magnetic properties can be fully defined in addition to the kinematic properties. We refer to this FR as FRED (Flux Rope from Eruption Data). We illustrate the FRED construction using the 2012 July 12 eruption and compare the coronal and interplanetary properties of the FR.  The results indicate that the FRED input should help make realistic predictions of the components of the FR magnetic field in the heliosphere. 

\keywords{coronal mass ejection, flux rope, reconnection.}
\end{abstract}

\firstsection 
\section{Introduction}

Coronal mass ejections (CMEs) are the most important players in space weather because they cause the severest of geomagnetic storms and accelerate energetic particles to GeV energies. The intensity of geomagnetic storms as measured by the $Dst$ index primarily depends on the magnitude of the south-pointing out-of-the-ecliptic ($B_z$) component of the interplanetary magnetic field and the speed ($V$) with which the structure impacts Earth's magnetosphere: $Dst$ = $-$0.01$V$$\mathopen|B_z\mathclose|$$-$32 nT (\cite[Wu \& Lepping 2002]{WuLepping02}; \cite[Gopalswamy et al. 2008]{Gopalswamy_etal08}). One of the vexing problems in space weather has been the prediction of southward (negative) $B_z$ of the interplanetary field that reconnects with Earth's magnetic field to produce geomagnetic storms. While there has been reasonable progress in the prediction of the arrival time of CMEs, $B_z$ prediction has been very limited due to the lack of realistic input to global MHD models that track CMEs into the heliosphere and provide asymptotic values of the CME parameters including the magnetic field. Since most CMEs arriving at Earth have a flux rope (FR) structure, it is inevitable that the global MHD models should use a FR input at the near-Sun boundary. Current models use a pressure pulse (which has no magnetic content) or ad hoc magnetic structures such as a spheromak (\cite[Odstrcil \& Pizzo 1999]{OdstrcilPizzo99}; \cite[Jin et al. 2017]{Jin_etal17}; \cite[Shen et al. 2014]{Shen_etal14}; \cite[Shiota et al. 2014]{Shiota_etal14}). \\ 
We construct a ``flux rope from eruption data" (FRED) by combining two key results: (i) the reconnected (RC) flux during an eruption approximately equals the poloidal flux of the ejected flux rope (Longcope et al. 2007; Qiu et al. 2007; Hu et al. 2014; Gopalswamy et al. 2017a), and (ii) white-light or EUV coronal mass ejections (CMEs) can be fit to a FR to get its geometrical properties (see e.g., Temmer et al. 2011). The RC flux is computed from the area under post-eruption arcades (PEAs, McAllister and Martin 2000) and the underlying unsigned photospheric magnetic field strength. The poloidal flux of the FR is known from the RC flux; assuming that the FR is force free (Lundquist 1951) we can get the axial and azimuthal field components and the toroidal flux of the flux rope. Thus we have a fully-defined FR, complete with geometric and magnetic parameters that can be used as input to global MHD models. The axial and poloidal fields are essentially responsible for geoeffectiveness in high and low-inclination magnetic clouds at Earth, respectively and hence the FR input should lead to definite prediction schemes. The FRED technique is complementary to another one that uses the source-region relative magnetic helicity along with FR geometrical properties to estimate the FR magnetic properties (Patsourakos et al. 2016).
\begin{figure}[t]
\begin{center}
 \includegraphics[width=4.6in]{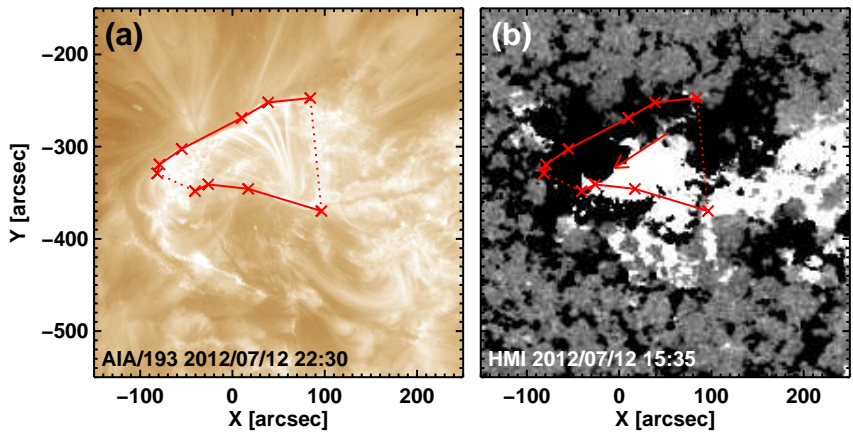} 
\caption{(a) PEA area (polygon) marked on an SDO/AIA 193 \AA\ image. (b) SDO/HMI line-of-sight magnetogram with PEA and the filament magnetic field direction (arrow) overlaid.}
   \label{fig1}
\end{center}
\end{figure}

\section{The Coronal Flux Rope}

We illustrate the construction of FRED using the 2012 July 12 eruption that resulted in a large SEP event and a major geomagnetic storm (Dst = -139 nT). The event has been studied by many authors (see e.g., \cite[Gopalswamy et al. 2013]{Gopalswamy_etal13}; \cite[Gopalswamy et al. 2014]{Gopalswamy_etal14}; \cite[Hess \& Zhang 2014]{HessZhang14}; \cite[M{\"o}stl et al. 2014]{M{\"o}stl_etal14}; \cite[Hu et al. 2016]{Hu_etal16}). Most of these papers were concerned with the kinematics and  Sun-to-Earth propagation of the CME that ended up as a shock-driving magnetic cloud at Earth. Many of the authors also fit a flux rope using the graduated cylindrical shell (GCS) model (\cite[Thernisien 2011]{Thernisien11}). Here we use the GCS model to get the geometric parameters of the FR. \\
{\underline{\it Geometrical Properties}}. Fitting the CME observed by the three views provided by the SOHO and STEREO missions to the GCS model, we get the coordinates of the flux rope as S12W06, which is slightly different from the flare location in AR 11520 (S15W01, \cite[Gopalswamy et al. 2014]{Gopalswamy_etal14}). The ratio of the FR radius ($R_0$) to the leading edge distance from the Sun center ($R_{tip}$) is 0.26. At $R_{tip}$ = 10 Rs, the $R_0$ = 2.6 Rs in the coronagraph field of view. The face-on and edge-on angular widths of the FR are 94$^{\circ}$ and 40$^{\circ}$, respectively. The tilt angle of the FR axis at its apex with respect to the horizontal is about 53$^{\circ}$, indicating a northwest-southeast orientation of the flux rope axis, consistent with the neutral line as inferred from filament location.  \\
{\underline{\it Magnetic Properties}}. We combine these geometrical information with the RC flux to get the magnetic properties of the FR.  It was recently shown that the total RC flux during an eruption can be obtained from a snapshot of the post eruption arcade (PEA) and the underlying photospheric magnetic field strength (\cite[Gopalswamy et al. 2017a]{Gopalswamy_etal17a}). Fig. 1 shows a snapshot of the PEA associated with the 2012 July 12 eruption as observed by the Atmospheric Imaging Assembly (AIA, Lemen et al. 2012) on board the Solar Dynamics Observatory (SDO, Pesnell et al. 2012) at 193 \AA. Also shown is the corresponding line-of-sight photospheric magnetic field from SDO’s Helioseismic and Magnetic Imager (HMI, Scherrer et al. 2012). The arrow in Fig. 1(b) points to the direction of the axial magnetic field of the erupted flux rope based on filament connectivity. The axis is tilted by about 46 degrees to the horizontal. The area of the polygon in Fig. 1(a) is 7.2$\times$10$^{19}$ cm$^2$, and an average magnetic field strength of $B$ of 392 G. The RC flux $\phi{_r}$ is thus 1.42$\times$10$^{22}$ Mx, which is half the unsigned flux through the PEA area.

 Since $\phi{_r}$ roughly equals the poloidal flux ($\phi{_p}$) of the erupted FR, we can get the axial field ($B_0$) of the coronal FR from the relation, $\phi{_p}$ = $(L/x_{01})B_0R_0$, where $L$ is the FR length and $x_{01}$ (=2.4048) is the first zero of the Bessel function $J_0$. Taking $L$ = 2 $R_{tip}$ = 20 Rs, we get $B_0$ = $\phi{_r}x_{01}/LR_0$ = 0.13 G, which is near the higher end of the axial field of coronal FRs reported in Gopalswamy et al. (2017b). The FR axial field strength is larger than the typical ambient field strength. Gopalswamy and Yashiro (2011) obtained the strength of the ambient magnetic field as $B_{amb}$ = 0.329$R$$^{-1.23}$, which gives $B_{amb}$ = 0.019 G at a heliocentric distance $R$ of 10 Rs. Clearly, the FR axial field strength is about 7 times greater than the ambient field, consistent with the fact that CME FRs are low-beta plasmas.  The toroidal flux of the FR is $\phi{_t}$ = $\phi{_p}$$(2\pi R_0/L)J_1(x_{01})$, where $J_1$ is the first order Bessel function. With the above numbers from the GCS fit, we get $\phi{_t}$ = 0.42$\phi{_p}$ or 5.96$\times$10$^{21}$ Mx. The poloidal ($B_p$) and toroidal ($B_t$) field strengths at any distance $r$ from the FR axis are given by $B_p$ = \begin{math}HB_0J_1(\alpha r)\end{math} with $H$ = $\pm1$ (helicity sign) and  $B_t$ = \begin{math} B_0J_{0}(\alpha r)\end{math}. Here, \begin{math} \alpha = x_{01}/R_0\end{math} is the force-free parameter. In the present case, $H$ = $+$1 because the axial field points mostly to the south and the azimuthal field goes from the positive to the negative side (see Fig. 1b). The direction is consistent with a southern hemispheric eruption having right-handed helicity sign (e.g., \cite[Bothmer \& Schwenn 1998]{BothmerSchwenn98}). Finally, the relative helicity per unit length ($H_r/L$) is also well-defined for a Lundquist FR: $H_r/L$ = 0.7$B_0$$^2$$R_0$$^3$ (e.g., Dasso et al. 2003). For the 2012 July 12 coronal FR, $H_r$ = 9.98 $\times$10$^{43}$ Mx$^2$, if we take the FR length to be 2$R_{tip}$.  Thus, the flux rope is fully defined both geometrically and magnetically. The three-dimensional CME speed was measured using STEREO-B as $V$ = 1548 km s$^{-1}$ (\cite[Gopalswamy et al. 2013]{Gopalswamy_etal13}). According to the empirical relation $V$ = 298$\phi{_r}$$^{0.75}$ (\cite[Gopalswamy et al. 2017b]{Gopalswamy_etal17b}), this speed indicates a $\phi{_r}$ of 9.0$\times$10$^{21}$ Mx, which is only about 36\% lower than the observed value (1.42$\times$10$^{22}$ Mx).
 
  \begin{figure}[]
\begin{center}
 \includegraphics[width=3. in]{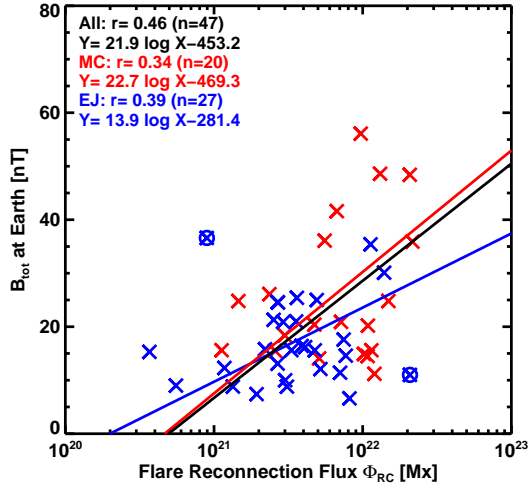} 
 \caption{Scatter plot between observed $B_{tot}$ and $\phi{_r}$ for a set of events from solar cycle 23 reported in Gopalswamy et al. (2017b). Two outliers are excluded from the correlation. Blue and red symbols correspond to magnetic clouds (MC) and non-cloud ejecta (EJ), respectively.}
   \label{fig2}
\end{center}
\end{figure}

\section{Comparison with 1-AU Observations}
A recent study showed that the axial field of the coronal FR derived from $\phi_{r}$ correlates with the axial field of the 1-AU flux rope (fitted) and also with the strength of the observed total magnetic field ($B_{tot}$)(\cite[Gopalswamy et al. 2017b]{Gopalswamy_etal17b}). In that study the white-light CME data were fit with the elliptical flux rope model of Krall and St. Cyr (2006). Figure 2 shows a scatter plot between $\phi{_r}$ and $B_{tot}$ for a set of interplanetary CME (ICME) events from cycle 23. From the regression line $B_{tot}$ = 21.9$\log$$\phi{_r}$$-$453.2 nT, we get $B_{tot}$ =31.9 nT with $\phi{_r}$ = 1.42$\times$10$^{22}$ Mx for the 2012 July 12 eruption. The inferred $B_{tot}$ is very close to the observed value at 1 AU: $\sim$30 nT (see e.g., Hess and Zhang 2014). If the FR expands self-similarly, the 1-AU FR radius is expected to be 0.26 AU (since $R_0$/$R_{tip}$ = 0.26 with $R_{tip}$ = 1 AU). Therefore, one can get the 1-AU axial field strength from the coronal field strength ($B_0$ = 0.13 G) as: $B_{01AU}$ = $B_0$(10/214)$^2$ = 28.3 nT, again very similar to the observed value.  The assumption of self-similar expansion seems to be valid because the FR radius is consistent with that obtained from Grad-Shafranov reconstruction using \textit{in-situ} data (Hu et al. 2016). For the same reasons, $H_r$ in the 1 AU FR is similar to that in the coronal FR. The 1-AU data also shows that the FR Y-component (azimuthal) of the magnetic field rotates from west to east, while the Z-component (axial) points southward throughout the cloud duration (Hess \& Zhang 2014; Hu et al. 2016), consistent with the orientation of the coronal FR. In this event, the $B_z$ is simply the axial field component.

\section{Discussion and Summary}
We have shown that the FRED technique is a viable starting point to infer the expected asymptotic magnetic structure in the heliosphere. It combines the line-of-sight photospheric magnetograms and PEA observations (X-ray, EUV, H-alpha, or microwave) with CME observations in the corona to obtain complete properties of coronal FRs. We have shown these properties at a distance of 10 Rs for the purpose of illustration, but FRED properties can be obtained at any initial height where the CME is observed. For example, the inner boundary of many MHD simulation models is located at $R_{tip}$ $\sim$ 21 Rs, where the axial field $B_0$ = $\phi{_r}x_{01}/LR_0$ will be weaker than the 10-Rs case because both $L$ and $R_0$ will be larger. Expressing $L$ and $R_0$ in terms of $R_{tip}$ and noting that $R_0$/$R_{tip}$ is invariant for each FR from the GCS model we get $B_0$ = $\phi{_r}x_{01}/[pq(R_{tip}$)$^2$], where $p$ and $q$ are constants: $L$ = $pR_{tip}$ (assumed FR length) and $R_0$ = $qR_{tip}$ (from the GCS model). This expression shows that the axial field falls off as the square of the heliocentric distance. Note that we have taken $p$ = 2 following Nindos et al. (2003). The length can be higher if the flux rope assumes the shape of a half torus, so $L$ = $\pi$$R_{tip}$. D{\'e}moulin et al. (2016) recently reported a statistical value of $p$ = 2.6$\pm0.3$, which is in between the two values noted above. The invariant toroidal flux can also be written in a general form as $\phi{_t}$ = $\phi{_p}$$(2\pi q/p)J_1(x_{01})$. Global MHD models that use a realistic input such as FRED should be able to provide a realistic forecast of what to expect in the heliosphere. It must be noted that additional effects such as CME deflection and rotation can modify the FR, which can be accounted for using semi-analytic models (see e.g., Kay et al. 2015). The results presented in this paper support the idea that CME FRs are formed during the eruption. The poloidal flux of any pre-existing FR is expected to be a small fraction of that added during the eruption. 

{\underline{\it Acknowledgement}}.
Work supported by NASA Heliophysics GI and LWS programs.


\begin{thebibliography}{}

\bibitem[Bothmer \& Schwenn(1998)]{1998AnGeo..16....1B} Bothmer, V., \& Schwenn, R.\ 1998, \textit{Annales Geophysicae}, 16, 1 

\bibitem[Dasso et al.(2003)]{2003JGRA..108.1362D} Dasso, S., Mandrini, C.~H., D{\'e}Moulin, P., \& Farrugia, C.~J.\ 2003, \textit{JGR}, 108, 1362 

\bibitem[D{\'e}moulin et al.(2016)]{2016SoPh..291..531D} D{\'e}moulin, P., Janvier, M., \& Dasso, S.\ 2016, \textit{SolPhys}, 291, 531 

\bibitem[Gopalswamy et al.(2008)]{2008JASTP..70..245G} Gopalswamy, N., Akiyama, S., Yashiro, S., Michalek, G., \& Lepping, R.\ 2008, \textit{JASTP}, 70, 245 

\bibitem[Gopalswamy \& Yashiro(2011)]{2011ApJ...736L..17G} Gopalswamy, N., \& Yashiro, S.\ 2011, \textit{ApJ}, 736, L17

\bibitem[Gopalswamy et al.(2013)]{2013SpWea..11..661G} Gopalswamy, N., M{\"a}kel{\"a}, P., Xie, H., \& Yashiro, S.\ 2013, \textit{Space Weather}, 11, 661

\bibitem[Gopalswamy et al.(2014)]{2014EP&S...66..104G} Gopalswamy, N., Xie, H., Akiyama, S., M{\"a}kel{\"a}, P.~A., \& Yashiro, S.\ 2014, \textit{Earth, Planets, and Space}, 66, 104 

\bibitem[Gopalswamy et al.(2017a)]{2017SoPh..292...65G} Gopalswamy, N., Yashiro, S., Akiyama, S., \& Xie, H.\ 2017a, \textit{SolPhys}, 292, 65 

\bibitem[Gopalswamy et al.(2017b)]{2017arXiv170508912G} Gopalswamy, N., Akiyama, S., Yashiro, S., \& Xie, H.\ 2017b, \textit{JASTP},  arXiv:1705.08912

\bibitem[Hess \& Zhang(2014)]{2014ApJ...792...49H} Hess, P., \& Zhang, J.\ 2014, \textit{ApJ}, 792, 49 

\bibitem[Hu et al.(2014)]{2014ApJ...793...53H} Hu, Q., Qiu, J., Dasgupta, B., Khare, A., \& Webb, G.~M.\ 2014, \textit{ApJ}, 793, 53 

\bibitem[Hu et al.(2016)]{2016ApJ...829...97H} Hu, H., Liu, Y.~D., Wang, R., M{\"o}stl, C., \& Yang, Z.\ 2016, \textit{ApJ}, 829, 97 

\bibitem[Jin et al.(2017)]{2017ApJ...834..173J} Jin, M., Manchester, W.~B., van der Holst, B., et al.\ 2017, \textit{ApJ}, 834, 173 

\bibitem[Kay et al.(2015)]{2015ApJ...805..168K} Kay, C., Opher, M., \& Evans, R.~M.\ 2015, \textit{ApJ}, 805, 168 

\bibitem[Krall \& St.~Cyr(2006)]{2006ApJ...652.1740K} Krall, J., \& St.~Cyr, O.~C.\ 2006, \textit{ApJ}, 652, 1740 

\bibitem[Lemen et al.(2012)]{2012SoPh..275...17L} Lemen, J.~R., Title, A.~M., Akin, D.~J., et al.\ 2012, \textit{SolPhys}, 275, 17 

\bibitem[Longcope et al.(2007)]{2007SoPh..244...45L} Longcope, D., Beveridge, C., Qiu, J., et al.\ 2007, \textit{SolPhys}, 244, 45 

\bibitem[Lundquist (1951)]{1951PhRv...83..307L} Lundquist, S.\ 1951, \textit{Phys. Rev.}, 83, 307

\bibitem[McAllister \& Martin (2000)]{2000AdSpR..26..469M} McAllister, H., Martin, S.~F.\ 2000, \textit{AdSpR}, 26, 469

\bibitem[M{\"o}stl et al.(2014)]{2014ApJ...787..119M} M{\"o}stl, C., Amla, K., Hall, J.~R., et al.\ 2014, \textit{ApJ}, 787, 119 

\bibitem[Nindos et al.(2003)]{2003ApJ...594.1033N} Nindos, A., Zhang, J., \& Zhang, H.\ 2003, \textit{ApJ}, 594, 1033 

\bibitem[Odstr{\v c}il \& Pizzo(1999)]{1999JGR...104..483O} Odstr{\v c}il, D., \& Pizzo, V.~J.\ 1999, \textit{JGR}, 104, 483

\bibitem[Patsourakos et al.(2016)]{2016ApJ...817...14P} Patsourakos, S., Georgoulis, M.~K., Vourlidas, A., et al.\ 2016, \textit{ApJ}, 817, 14 

\bibitem[Pesnell et al.(2012)]{2012SoPh..275....3P} Pesnell, W.~D., Thompson, B.~J., \& Chamberlin, P.~C.\ 2012, \textit{SolPhys}, 275, 3

\bibitem[Qiu et al.(2007)]{2007ApJ...659..758Q} Qiu, J., Hu, Q., Howard, T.~A., \& Yurchyshyn, V.~B.\ 2007,\textit{ApJ}, 659, 758 

\bibitem[Scherrer et al.(2012)]{2012SoPh..275..207S} Scherrer, P.~H., Schou, J., Bush, R.~I., et al.\ 2012, \textit{SolPhys}, 275, 207 

\bibitem[Shen et al.(2014)]{2014JGRA..119.7128S} Shen, F., Shen, C., Zhang, J., et al.\ 2014, \textit{JGR}, 119, 7128 

\bibitem[Shiota et al.(2014)]{2014SpWea..12..187S} Shiota, D., Kataoka, R., Miyoshi, Y., et al.\ 2014, \textit{Space Weather}, 12, 187 

\bibitem[Temmer et al.(2011)]{2011SoPh..273..421T} Temmer, M., Veronig, A.~M., Gopalswamy, N., \& Yashiro, S.\ 2011, \textit{SolPhys}, 273, 421 

\bibitem[Thernisien(2011)]{2011ApJS..194...33T} Thernisien, A.\ 2011, \textit{ApJS}, 194, 33 

\bibitem[Wu \& Lepping(2002)]{2002JGRA..107.1346W} Wu, C.-C., \& Lepping, R.~P.\ 2002, \textit{JGR}, 107, 1346 

\end{thebibliography}
\end{document}